\documentclass[letterpaper]{article} 
\usepackage{aaai2026}  
\nocopyright
\usepackage{times}  
\usepackage{helvet}  
\usepackage{courier}  
\usepackage[hyphens]{url}  
\usepackage{graphicx} 
\usepackage{natbib}  
\usepackage{caption} 
\usepackage{algorithm}
\usepackage{algorithmic}

\usepackage{newfloat}
\usepackage{listings}
\DeclareCaptionStyle{ruled}{labelfont=normalfont,labelsep=colon,strut=off} 
\floatstyle{ruled}
\newfloat{listing}{tb}{lst}{}
\floatname{listing}{Listing}

\usepackage{xcolor}
\definecolor{efficiency}{HTML}{E97132}
\definecolor{practicality}{HTML}{0F9ED5}
\definecolor{integrity}{HTML}{4EA72E}
\definecolor{nexbaxbox}{HTML}{F2F2F2}

\usepackage{tabularx}
\newcommand{\midtilde}{\raisebox{-0.5ex}{\textasciitilde}}

\usepackage{enumitem}
\title{Next-Billion AI Index: The compass for AI utility and adoption\\in the global majority}
\author{
    Ambrish Rawat\textsuperscript{\rm 1},
    Jessica He\textsuperscript{\rm 1},
    Subhabrata Majumdar\textsuperscript{\rm 2},
    Claudio Pinhanez\textsuperscript{\rm 3},
    Yann Le Beux\textsuperscript{\rm 4},
    Satyapriya Krishna\textsuperscript{\rm 5},
    Rahul Gupta\textsuperscript{\rm 6},
    Rumman Chowdhury\textsuperscript{\rm 7},
    Kush R. Varshney\textsuperscript{\rm 1}
}
\affiliations{
    \textsuperscript{\rm 1}IBM Research \textsuperscript{\rm 2}Indian Institute of Management Bangalore \textsuperscript{\rm 3}University of São Paulo \textsuperscript{\rm 4}YUX Design\\ \textsuperscript{\rm 5}Harvard University \textsuperscript{\rm 6}Amazon \textsuperscript{\rm 7}Humane Intelligence\\

}

\begin{document}

\maketitle

\begin{abstract}
Generative AI assessments remain dominated by frontier capability benchmarks that often fail to capture whether systems can be sustainably deployed, adapted, and trusted in locally grounded and infrastructure-constrained settings.
This paper introduces the Next Billion AI Index (\emph{nexbax}), which we believe is the first diagnostic framework to treat economic viability, operational deployability, and governance alignment as co-equal determinants of AI utility in next-billion-user contexts.
Rather than treating usefulness as a single outcome, \emph{nexbax} operationalizes the preconditions for useful AI through 10 dimensions organized under three themes: Effective Efficiency, Operational Practicality, and Societal Integrity.
These dimensions assess whether systems are economically viable, deployable under infrastructure and workflow constraints, and aligned with local needs, user expectations, and collaborative development practices.
We pair the framework with rubrics for weak, moderate, and strong performance, and conduct a formative expert evaluation through eleven semi-structured interviews with founders, developers, product leaders, and technical practitioners building AI systems for next-billion markets.
Participants found the index useful for reasoning about adoption trade-offs and effective at capturing factors shaping real-world AI uptake---particularly cost, usability, reliability, and trust. 
They also identified the need for contextual explanations, domain-specific evidence, and broader stakeholder validation.
\emph{Nexbax} is therefore proposed not as a universal score of social value, but as a diagnostic for artificial \emph{useful} intelligence: a way to make visible the technical, economic, and governance properties that make inclusive AI deployment more viable.
\end{abstract}


\section{Introduction}

What determines whether AI meaningfully benefits the next billion users? While AI has shown significant promise to multiply user productivity in western, technological contexts, it is yet to replicate that promise for users in the global majority.
In next-billion settings characterized by people using digital technologies for the first time in emerging markets, artificial intelligence (AI) adoption is limited by fundamental constraints: meaningful connectivity and affordability, mobile-first device realities, optimized compute and energy reliability, fragmented and informal economic activity, and heterogeneous language and literacy contexts~\cite{oluwatuyi2026collectively, varuvel2026designing}. Under these conditions, the primary determinant of real-world AI impact is not peak benchmark performance, but whether systems can be deployed and sustained under local constraints \cite{UNDP2025}. This stands in sharp contrast to high-income markets, where AI diffusion is increasingly shaped by enterprise software integration, abundant cloud compute, mature data pipelines, and relatively affordable connectivity, allowing ``model capability'' to translate more quickly into productivity gains. In line with this mismatch, interviews with practitioners deploying AI in the Global South found that dominant AI evaluation benchmarks were insufficient for local contexts, and instead, ``evaluation practices had to evolve toward participatory, context-sensitive approaches that captured nuances generic benchmarks could not represent''~\cite{varuvel2026designing}. Cross-country evidence also suggests that the character of adoption differs across these contexts. In University of Melbourne’s and KPMG’s 2025 ``Trust, Attitudes and Use of Artificial Intelligence'' global study \cite{KPMG2025}, which surveyed over 48,000 respondents in 47 countries, people in emerging economies such as India, Egypt, and Nigeria reported markedly higher trust in AI systems than those in high-income economies, with emerging economies showing \midtilde57\% trust versus \midtilde39\%. Likewise, Anthropic’s Economic Index \cite{Anthropic2026} and Microsoft’s 2025 Future of Work report \cite{Microsoft2025} suggest that, whereas AI use in high-income economies is concentrated in white-collar knowledge work and enterprise workflow automation, in emerging markets it more often functions as a general-purpose problem-solver for informal work, small businesses, education, and frontline services, where flexibility, cost efficiency, and usability matter more than frontier model performance~\cite{belcak2025small}. Taken together, these patterns underscore the need for evaluative frameworks that assess AI not by abstract model capability or aspirations toward artificial general intelligence, but by what might be termed artificial \emph{useful} intelligence: economic viability and contextual utility under real-world constraints.

India and markets across Africa represent large and diverse next-billion-user contexts that make these dynamics concrete.

In India, the ``next billion AI market'' is not a monolithic consumer segment but a digitally amplified, institutionally mediated adoption environment in which AI value is expected to scale through public platforms, regulated sectors, and micro, small and medium enterprise participation. The government of India's policy think tank positions AI as a macroeconomic lever, but it also makes clear that readiness depends on more than model capability: it depends on shared digital rails, compute access, interoperable datasets, sector-grade governance, and domain-linked skilling \cite{NITIAayog2025}. As a result, product-market fit in India privileges affordability, robustness, multilingual access, and compliance-by-design, while foregrounding actors such as digitally powered infrastructure (DPI) custodians, regulators, and sector-specific developers as key market makers shaping what can be deployed, at what cost, and for whom.

Across Africa, the ``next billion AI market'' is better understood as a people-centered, development-oriented adoption environment where value is defined less by frontier capability than by practical utility under binding constraints. Participatory research with members of a marginalized community in South Africa found a strong desire to leverage AI for socioeconomic advancement \cite{oluwatuyi2026collectively}. Similarly, the African Union’s Continental AI Strategy \cite{AU2024} frames AI as an engine for inclusive growth and treats governments as anchor adopters that can catalyze deployment through DPI, open data, procurement, and partnerships. At the same time, the strategy highlights that uptake depends on closing foundational gaps in electricity, broadband, data infrastructure, compute access, and locally relevant datasets, while treating governance and risk management as first-order adoption conditions. This vision is operationalized through targeted national missions, such as Nigeria’s National Artificial Intelligence Strategy (2025–2029), which explicitly links AI infrastructure to local talent development and social inclusion \cite{NigeriaNAIS2025} --- and Kenya’s National AI Strategy 2025–2030 which underscores the use of community-based `Digital Innovation Hubs' to ensure AI technologies remain accessible to underserved groups and local SMEs \cite{KenyaAI2025}. In this context, systems are likely to be rewarded for being cost- and energy-efficient, multilingual and locally grounded, deployable through public and SME channels, and governed in ways that sustain trust.

Together, India and African markets illustrate a shared structural reality: AI adoption in next-billion contexts is determined less by frontier performance and more by the interplay of affordability, infrastructural variance, linguistic diversity, institutional capacity, and the centrality of public and small-enterprise channels. A related theme is ``sovereign AI'': the ability of states, firms, and communities to retain legitimate control and agency across the AI stack, rather than becoming dependent on monopolistic providers at critical layers of data, compute, models, or deployment infrastructure \cite{Roberts2024DigitalSovereignty}. These determinants sit largely outside prevailing, capability-centric benchmarks \cite{METRAutonomousCapabilities,METRTimeHorizons,liang2023holistic}, which implicitly assume mature infrastructure and enterprise capacity. In such settings, ``AI progress'' cannot be equated with benchmark supremacy; it must be defined by economic viability, operational resilience, and contextual utility under real-world constraints.

The Next Billion AI Index (\emph{nexbax}) addresses this measurement blind spot. It evaluates generative AI systems---including models and agentic architectures---not by abstract performance metrics alone, but by whether they exhibit system-level properties that can support economically sustainable, inclusive, and developmentally relevant deployment across constrained environments.
Inspired by the precedent set by Stanford’s Foundation Model Transparency Index (FMTI)~\cite{FMTI2023}, which spurred greater openness across the AI industry, \emph{nexbax} extends that momentum toward inclusivity, accessibility, and innovation for the next billion users, asking not only how powerful systems are, but whether they create the technical, economic, and governance conditions under which meaningful use becomes more likely. Partly informed by earlier frameworks for inclusive innovation, including C. K. Prahalad's work on technology innovation for the next billion \cite{Prahalad2006,Prahalad2012}, \emph{nexbax} reinterprets these concerns for the generative AI era through \textbf{10 measurable dimensions} organized under \textbf{3 themes}: \textbf{\textcolor{efficiency}{Effective Efficiency}}, \textbf{\textcolor{practicality}{Operational Practicality}}, and \textbf{\textcolor{integrity}{Societal Integrity}}. These capture how well AI solutions balance affordability, adaptability, and social responsibility. In this version, we pair the dimensions with broad rubrics and a formative expert exercise to explore how such a diagnostic can support comparison across AI technologies. In doing so, it aims to \textbf{incentivize innovation that is not only powerful but also socioeconomically viable}, while recognizing that actual usefulness and cultural appropriateness must be established through situated evaluation with local stakeholders and affected communities.
\begin{center}
\fcolorbox{black!20}{nexbaxbox}{%
\begin{minipage}{0.92\linewidth}
\emph{Nexbax} is not a universal measure of usefulness for the global majority. It is a first-stage diagnostic for identifying whether AI systems possess the pre-conditions---technical, economic, and governance properties---likely to support useful deployment for the next-billion. Determining actual usefulness, cultural appropriateness, and social value requires situated evaluation with local stakeholders and affected communities.
\end{minipage}}
\end{center}

The remainder of the paper situates \emph{nexbax} in related work on inclusive innovation and AI evaluation (Section~\ref{sec:related_work}), defines the index dimensions and rubrics (Section~\ref{sec:nexbax}), reports findings from a formative expert evaluation (Section~\ref{sec:evals}), discusses limitations and future directions for participatory, context-specific validation (Section~\ref{sec:limitations}), and concludes by reflecting on the broader implications of evaluating AI through the lens of useful and inclusive adoption (Section~\ref{sec:conclusions}).

\section{Related Work}
\label{sec:related_work}

\subsection{Influence of AI Indices}

Practically, indices play a crucial role in aligning innovation incentives by translating complex, multidimensional goals into comparable and trackable signals. In the AI ecosystem, FMTI \cite{FMTI2023} demonstrated how clearly defined indicators can shift industry behavior by elevating transparency from a normative aspiration to a measurable performance dimension. \emph{Nexbax} builds on this logic by extending evaluative attention beyond transparency toward the conditions that enable broad-based adoption and value creation. By benchmarking AI systems across dimensions that matter for low-resource and underrepresented contexts, the index helps redirect optimization efforts toward efficiency, interoperability, and user empowerment areas that are often underincentivized despite their outsized impact on real world uptake~\cite{Rismani_Shelby_Davis_Rostamzadeh_Moon_2025}.

Crucially, \emph{nexbax} functions as a diagnostic instrument for equitable development rather than a prescriptive ranking of technological superiority. By making gaps in affordability, resilience, localization, and openness empirically visible, the index enables developers, investors, and policymakers to identify where targeted innovation can unlock disproportionate social and economic value. Over time, tracking through a standardized index allows stakeholders to distinguish between nominal advances in AI capability and substantive progress in inclusive utility. In this way, \emph{nexbax} supports a development-aligned innovation trajectory, one in which success is measured not only by what AI systems can do in ideal settings, but by how effectively they serve diverse populations at scale.

\subsection{Related Indices and Frameworks}

This subsection reviews adjacent efforts that inform \emph{nexbax} while also clarifying the gap it is intended to fill. The most relevant literature spans AI-wide indices and comparative frameworks; national missions and public-sector assessment efforts; accountability and transparency frameworks; domain-specific application literatures; cross-cutting capability verticals such as multilinguality, efficiency, and robustness; and broader traditions of inclusive innovation. Taken together, these streams show that AI progress is already being measured from multiple angles, but they rarely integrate deployability, inclusivity, governance alignment, and developmental relevance into a single comparative framework for next-billion-user contexts.

Conceptually, \emph{nexbax} is informed by value-sensitive design \cite{BorningMuller2012}, which emphasizes embedding human values such as accessibility, fairness, and accountability directly into technical systems rather than treating them as downstream constraints. By operationalizing these values as measurable system-level properties, the index moves beyond aspirational principles toward actionable signals that can shape design and optimization choices.

\subsubsection{AI indices and benchmarks.}
A growing body of work has introduced indices and benchmarking frameworks to measure different aspects of AI development, diffusion, and governance. Early and influential efforts such as the Stanford AI Index \cite{AIIndex2017} provide ecosystem-level trend tracking by aggregating indicators across research output, model capabilities, investment flows, and deployment patterns, offering a longitudinal view of the ``state of AI''. Similar macro-level indices and readiness assessments—developed by institutions such as the OECD.AI index \cite{OECD2026} and Oxford Insights’ Government AI Readiness index \cite{Oxford2025}—evaluate national or regional capacity for AI adoption, emphasizing infrastructure, talent pipelines, regulatory frameworks, and public-sector readiness. Complementing these, the Global Index on Responsible AI (GIRAI) by the Global Center on AI Governance based in South Africa provides a comparative global benchmark specifically focused on the implementation of responsible AI commitments across diverse national contexts \cite{GIRAI2024}. Capability-focused efforts such as METR's autonomous-agent evaluations and time-horizon studies, along with HELM's holistic language-model evaluation framework, provide more system-level evidence about model performance and emerging autonomous capabilities \cite{METRAutonomousCapabilities,METRTimeHorizons,liang2023holistic}. While these indices and benchmarks are valuable for comparability and agenda-setting, they are primarily descriptive or capability-centric, and 
provide limited insight into how individual AI systems perform across heterogeneous, culturally diverse, and resource-constrained contexts, or how their downstream societal impacts should be evaluated. Emerging initiatives are beginning to address these gaps by foregrounding cultural context, participatory design, and societal impact in AI evaluation and deployment. For example, Microsoft Research Africa’s Atlas: A Playbook for Cross-Cultural AI emphasizes culturally grounded approaches to AI system design and deployment across diverse sociotechnical settings \cite{atlas2025}, while the 
ImpactBench initiative seeks to develop benchmarks for evaluating the societal impacts of AI systems beyond conventional capability metrics \cite{ImpactBench2026}.

\subsubsection{National and multilateral AI missions.}
An adjacent but distinct category of measurement has emerged through large-scale national AI missions and public-sector programmes, particularly in countries with large populations and diverse socio-economic conditions. Initiatives such as the IndiaAI Mission \cite{IndiaAI}, as well as continental and regional strategies articulated through bodies such as the African Union \cite{AU2024}, increasingly incorporate scorecards, dashboards, or progress indicators to track investments, skilling efforts, startup ecosystems, and public-sector AI deployment. While these efforts are critical for coordinating state capacity and signaling national priorities, their evaluative focus is typically top-down and policy-centric, centered on programme execution rather than on the technical and operational characteristics of AI systems themselves. In contrast, \emph{nexbax} is intentionally bottom-up and system-oriented: it does not assess national readiness or mission success, but instead evaluates whether the AI technologies emerging within and often deployed across such ecosystems are practically usable, affordable, and adaptable for large, diverse user bases.

\subsubsection{Accountability, transparency, governance.}
A second line of work focuses on accountability, transparency, and responsible AI practices. The Stanford's FMTI \cite{FMTI2023} demonstrates how structured rubrics can operationalize transparency and incentivize improved disclosure by AI developers across data, evaluation, compute, and governance dimensions. Related assessment frameworks, including the AI Policy Observatory of the OECD \cite{OECD2026}, the Readiness Assessment Methodology for AI proposed by UNESCO \cite{UNESCO2023}, and ISO/IEC 42001's AI management-system standard \cite{ISOIEC420012023}, similarly emphasize institutional practices, documentation quality, and ethical alignment. These approaches have proven effective as soft-governance mechanisms; however, they largely evaluate organizational processes and reporting practices rather than downstream properties such as affordability, deployability, usability, or robustness of AI systems in low-infrastructure or culturally diverse settings~\cite{kim2025ai}.

\subsubsection{Application verticals.}
Another important stream of related work comes from domain-specific AI and digital system deployments in sectors such as healthcare, education, environment, agriculture, finance, and crisis response \cite{varuvel2026designing}. This literature is typically solution-oriented, emphasizing task-specificity and pragmatic adaptations such as offline-first architectures, low-bandwidth interfaces, voice and SMS-based interaction, support for low-literacy users, and compatibility with low-end devices to enable operation under constraints of infrastructure, capital, and institutional capacity. Collectively, this literature shows that meaningful functionality in resource-constrained settings depends less on maximal technical sophistication than on fit with local realities.

At the same time, this body of work remains highly contextual. Benchmarks or evaluations developed for one vertical often privilege domain-specific outcome measures, data conditions, and workflow assumptions, making them difficult to compare across sectors or to generalize into a common framework for AI system design. \emph{Nexbax} builds on the practical lessons surfaced in these sectors, but abstracts them into cross-cutting dimensions that can be applied across model families, developer tools, and deployment settings.

\subsubsection{Capability verticals.}
Parallel to application-specific benchmarking, a growing set of studies evaluates cross-cutting AI capabilities that are especially salient in next-billion-user contexts, including multilingual performance, low-resource language support, efficiency under compute and bandwidth constraints, robustness to noisy inputs, and interoperability across tools and platforms \cite{bailey2026street, microsoft2024vibhasha, muchai2026paza, pazabench2026, oluwatuyi2026collectively, varuvel2026designing}. These efforts are valuable because they expose failure modes that broad benchmark suites often miss, particularly where users, developers, and institutions operate outside high-resource English-centric settings.

However, capability-vertical assessments usually isolate one attribute at a time. A multilingual benchmark may reveal language inequities without speaking to affordability; an efficiency benchmark may highlight compute trade-offs without addressing transparency, governance, or user empowerment~\cite{miranda2026multilinguality}. For \emph{nexbax}, these verticals are therefore best understood as necessary but partial inputs into a broader evaluative frame that asks how these properties interact in real deployment environments.

Taken together, the literatures above provide many of the ingredients that \emph{nexbax} seeks to integrate: macro-level AI tracking, mission-driven public adoption goals, governance-oriented disclosure frameworks, lessons from sectoral deployments, and focused assessments of specific capabilities. What remains less developed is a unifying framework for comparing AI systems according to whether they are economically viable, operationally practical, and socially aligned for large, diverse, and resource-optimized settings. To motivate that synthesis, we next turn to broader frameworks for inclusive innovation on which \emph{nexbax} builds.

\subsubsection{Inclusive innovation.}
Prahalad’s framework provides an early and influential articulation of how technologies must be rethought to serve large, underserved populations  \cite{Prahalad2012,Prahalad2006}. Many of its core principles—such as radical price-performance improvements, robustness to hostile environments, scalability across cultures and languages, and redesign from first principles—remain relevant for AI systems intended for next-billion-user contexts. In particular, requirements around affordability, resilience, adaptability of user interfaces, and functionality under infrastructural volatility closely parallel challenges faced by contemporary AI deployments in low-resource settings.

More recent work on inclusive and globally situated AI has extended these concerns by emphasizing structural inequality, local participation, representational harms, and the need to account for diverse social and cultural contexts in AI design and governance \cite{Mohamed2020,Sambasivan2021,Prabhakaran2022,f2025exploring}. This literature highlights that inclusion cannot be reduced to access alone, but also depends on whose values, languages, institutional assumptions, and lived experiences are reflected in AI systems.

However, the broader inclusive innovation tradition has also been extensively critiqued for overemphasizing provider perspectives, market opportunity, and the framing of low-income populations primarily as consumers. Karnani, for example, argues that poverty alleviation requires attention to people as producers and workers, not only as buyers of low-cost products \cite{Karnani2007}. More broadly, the experimental development economics tradition associated with Banerjee, Duflo, and Kremer emphasizes that claims about poverty alleviation should be grounded in specific, empirically evaluated interventions among the people most affected, rather than inferred from broad market logics \cite{BanerjeeDuflo2009,banerjee2026poor}. These critiques are directly relevant to AI evaluation: a system can be affordable, technically robust, or attractive to providers while still failing to be useful, appropriate, or empowering in a particular community. \emph{Nexbax} therefore treats Prahalad's principles as one historical input rather than as a sufficient foundation.

\section{\emph{Nexbax}: Next Billion AI Index}
\label{sec:nexbax}

Building on these indices and frameworks, we developed the initial version of \emph{nexbax} using an iterative process, then validated and refined it through a formative user study. We describe the iterative process below and the study in Section~\ref{sec:evals}.

\subsection{Development of \emph{nexbax}}
We began by reviewing existing literature on digital inclusion, identified through searches for ``inclusive innovation,’’ ``digital inclusion,’’ ``open source AI''and ``AI adoption’’~\citep{belcak2025small, kim2025ai, kirk2024prism, melville2013amplifying, nature2025localizing, odame-darkwah2025isf, Prahalad2006,varshney2023foundation, seger2023open, casper2025open, eiras2024risks}. We mapped and reinterpreted the principles outlined in this literature in the context of AI systems using documentation of two leading closed-source and open-sourced models as guidance: GPT~\cite{achiam2023gpt}, Claude~\cite{anthropic2024claude3}, Mistral~\citep{jiang2023mistral, jiang2024mixtral}, Llama~\cite{touvron2023llama}, and Nemotron~\cite{chandiramani2026nemotron,adler2024nemotron}. Within their documentation, we identified instances of existing principles and adjacent concepts, then examined how their meanings and implications changed in the context of AI development and deployment. For example, \citet{Prahalad2006}'s principle of designing for ``hostile environments’’ manifested as mechanisms for ``robustness’’ in constrained environments, and ``deskilling’’ was reinterpreted as ``usability’’ of installation and deployment processes for AI systems.

Furthermore, to ensure relevance to diverse stakeholders --- including end users, developers, business decision-makers, and policy-makers --- one author held informal conversations based around the empathy map framework \cite{gibbons2018empathy} with 5 people from these stakeholder groups. Their feedback and priorities further informed the development of AI-specific considerations for global adoption.

Once an initial set of dimensions had been drafted, we referenced industry reports, such as the India AI Impact Summit goals \cite{indiaai2026impact} and the the ``Trust, Attitudes and Use of Artificial Intelligence'' global study \cite{KPMG2025}, to prioritize and refine components of our framework based on considerations grounded in real-world practice. For example, we removed sustainability as it primarily reflected infrastructural concerns rather than adoption priorities, and validated ``adaptability’’ and ``multiculturalism’’ due to their prevalence in such reports.

\begin{figure}
    \centering
    \includegraphics[width=\linewidth]{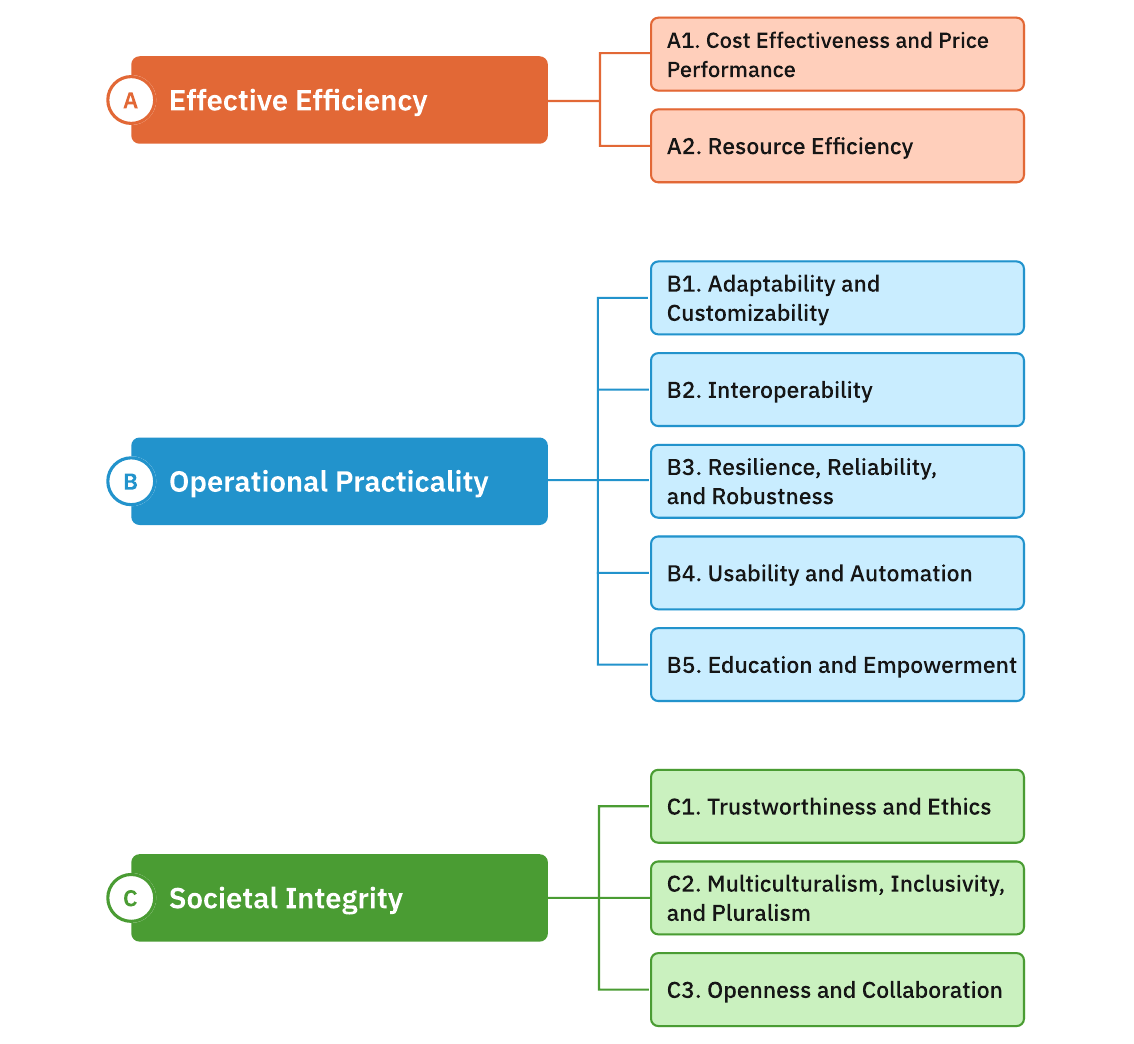}
    \caption{Overview of the Next Billion AI Index dimensions.}
    \label{fig:nexbax-overview}
\end{figure}

Finally, the leading author clustered the dimensions into three themes to reduce fragmentation and promote conceptual clarity. The resulting index is comprised of ten dimensions under three overarching themes (Figure~\ref{fig:nexbax-overview}) with concrete evaluation criteria for each dimension (Table~\ref{tab:nexbax-rubrics}). 



\subsection{Themes and Dimensions}
The \emph{nexbax} dimensions are organized into three themes that reflect where inclusivity is negotiated in modern AI systems: economic feasibility, operational deployability, and societal alignment. This thematic grouping is intentional. Rather than treating affordability, usability, and responsibility as independent attributes, \emph{nexbax} recognizes them as interacting layers of system design that jointly determine whether generative AI can be meaningfully adopted at scale in next-billion-user contexts. The three themes also align with distinct decision horizons such as cost and resource choices made early, deployment and adaptation challenges faced during rollout, and longer-term social and institutional consequences, allowing the index to surface trade-offs that are often obscured in capability-centric benchmarks. 

\subsubsection{Theme A: Effective Efficiency}
\emph{Why economic efficiency must be foundational, not downstream.}

Effective Efficiency captures whether an AI system can deliver acceptable utility under binding economic and infrastructural constraints. In global-majority contexts, affordability is not a secondary optimization objective but a gating condition: systems that are performant yet economically misaligned are functionally inaccessible. \emph{Nexbax} therefore treats cost-effectiveness and resource efficiency as first-order evaluative dimensions, reflecting both monetary cost (pricing models, total cost of ownership) and non-monetary resource intensity (compute, energy, bandwidth). 


This theme goes beyond nominal “cheapness.” It evaluates whether efficiency is structurally encoded, for example, through model size choices, edge or offline capability, and pricing flexibility, rather than achieved through temporary subsidies or scale advantages. Table~\ref{tab:measurement_signals} offers indicative signals for assessing these dimensions across AI systems. 

\subsubsection{Theme B: Operational Practicality}
\emph{Why deployability, not capability, determines real-world impact.}

Operational Practicality assesses whether AI systems can be adapted, maintained, and evolved in heterogeneous and often fragile environments. While many AI benchmarks implicitly assume stable connectivity, skilled operators, and homogeneous users, \emph{nexbax} foregrounds the frictions that dominate real deployments: intermittent infrastructure, linguistic diversity, limited developer bandwidth, and evolving use cases. 

The dimensions grouped here---adaptability, interoperability, robustness, usability/automation, and education/empowerment---collectively capture the translation layer between raw AI capability and sustained use. Importantly, several of these dimensions are developer-facing rather than end-user-facing, reflecting the reality that inclusivity is often decided upstream through tooling, abstractions, and integration pathways. Table~\ref{tab:measurement_signals} summarizes possible signals for these properties without privileging any single deployment paradigm. 

\subsubsection{Theme C: Societal Integrity}
\emph{Why inclusive AI requires shared stewardship, not just safeguards.}

Societal Integrity captures whether AI systems are aligned with the values, institutions, and pluralism of the populations they serve. Rather than treating ethics, inclusivity, and openness as compliance checklists, \emph{nexbax} frames them as enablers of legitimacy and long-term adoption. Systems that are opaque, culturally narrow, or closed to local participation and ownership may achieve short-term deployment but struggle to earn trust or evolve responsibly. 

The dimensions in this theme assess whether AI systems support multicultural representation, transparent governance, and collaborative development models that allow diverse stakeholders- developers, public institutions, and communities—to meaningfully shape outcomes. As suggested in Table~\ref{tab:measurement_signals}, these dimensions emphasize observable design and governance choices over aspirational commitments. 

\subsection{Measurement Signals}

To move from conceptual dimensions to evaluable properties, \emph{nexbax} treats each dimension as a composite construct that can be examined through observable signals. In this first version, these signals are intended as a practical starting point rather than a fixed scoring protocol. They combine standardized benchmarks where available with system-level and deployment-level evidence such as documentation, deployment requirements, pricing structures, language support, and integration pathways.

Table~\ref{tab:measurement_signals} situates the \emph{nexbax} dimensions within the current measurement landscape. It is indicative rather than exhaustive: mature benchmarks can inform some dimensions, while others require proxy-based evidence and situated human interpretation.
\begin{table*}[t]
\centering
\small
\begin{tabular}{p{0.18\textwidth} p{0.24\textwidth} p{0.24\textwidth} p{0.24\textwidth}}
\hline
\textbf{\emph{Nexbax} dimension} & \textbf{Measurement direction} & \textbf{Existing benchmarks / signals} & \textbf{Measurement gap} \\
\hline

\textbf{\textcolor{efficiency}{Cost-effectiveness and Price-performance}} & Utility per dollar; cost per successful task; total cost of ownership; availability of low-cost or community-access modes & Pricing primitives/disclosures; subscription tiers; efficiency-aware evaluation suites such as HELM~\cite{liang2023holistic}; deployment cost estimates & No widely adopted benchmark links task utility, local affordability, and long-term cost of ownership \\

\textbf{\textcolor{efficiency}{Resource efficiency}} &
Utility per watt; tokens or outputs per joule; latency; memory footprint; bandwidth requirements; offline or edge viability & MLPerf~\cite{reddi2020mlperf}Inference; AI Energy Score~\cite{AIEnergyScore2026}; Intelligence per watt~\cite{saad2025intelligence}; hardware and deployment footprint disclosures &
Energy and compute benchmarks rarely capture low-connectivity, mobile-first, or intermittent-infrastructure deployment conditions \\

\textbf{\textcolor{practicality}{Adaptability and \newline Customizability}} &
Fine-tuning support; adapter or LoRA support; prompt and configuration flexibility; domain adaptation; multimodal extensibility &
Evidence from model documentation, fine-tuning APIs, adapter ecosystems, and deployment templates, tools/agent-stack &
No standardized benchmark captures how easily systems can be adapted to local domains, workflows, or constraints \\

\textbf{\textcolor{practicality}{Interoperability}} &
Standards support; API and SDK availability; connectors; data portability; ease of integration with existing tools and workflows &
OpenAPI-compatible APIs; SDKs; platform integrations (LangChain/LlamaIndex/Ollama); vector database connectors; webhook and export support &
Existing signals are ecosystem-specific and do not provide a unified interoperability score \\

\textbf{\textcolor{practicality}{Resilience, Reliability, and Robustness}} &
Performance under noisy inputs, poor connectivity, outages, out-of-distribution prompts, dialectal variation, and adversarial behavior &
HELM robustness evaluations; DecodingTrust~\cite{wang2023decodingtrust}; stress tests for safety, bias, privacy, and adversarial behavior; red-teaming; bug bounty programs; API reliability/graceful degradation&
Lab robustness tests only partially reflect hostile or resource-constrained deployment environments \\

\textbf{\textcolor{practicality}{Usability and \newline Automation}} &
Setup time; number of steps to first working deployment; low-code/no-code support; agentic workflow reliability; documentation clarity &
Developer benchmarks such as SWE-Bench~\cite{jimenez2024swe} and BigCodeBench~\cite{zhuo2025bigcodebench}; onboarding flows; deployment tutorials; human task-completion studies &
Usability is highly context-dependent and often requires practitioner judgment rather than automated scoring alone \\

\textbf{\textcolor{practicality}{Education and \newline Empowerment}} &
Quality of documentation; tutorials; local-language help; training materials; community support; examples for low-resource settings &
Documentation audits; community forum activity; localized tutorials; public training programs; NGO or public-sector enablement materials &
Few mature benchmarks assess whether users or developers are empowered to understand, adapt, and maintain systems \\

\textbf{\textcolor{integrity}{Trustworthiness \newline and Ethics}} &
Safety behavior; bias evaluation; privacy protections; transparency artifacts; user controls; incident reporting channels &
MLCommons AILuminate~\cite{ghosh2025ailuminate}; DecodingTrust~\cite{wang2023decodingtrust}; model cards; system cards; safety and bias evaluation reports &
General safety benchmarks may miss context-specific harms, local norms, and institutional trust requirements \\

\textbf{\textcolor{integrity}{Multiculturalism, \newline Inclusivity, and \newline Pluralism}} &
Language and dialect coverage; low-resource language performance; cultural appropriateness; low-literacy interaction; locale-sensitive UX &
FLORES-200~\cite{goyal2022flores}; MTEB/MMTEB~\cite{muennighoff2023mteb,enevoldsen2025mmteb}; multilingual and low-resource language benchmarks; localization documentation &
Language benchmarks do not fully capture cultural fit, pluralism, dialectal nuance, or accessibility for low-literacy users \\

\textbf{\textcolor{integrity}{Openness and \newline Collaboration}} &
Open weights, code, and data; licensing; reproducibility; public roadmap; community governance; contribution pathways &
Open-weight releases; open-source repositories; license terms; issue trackers; model cards; public roadmaps &
Openness is multidimensional and cannot be inferred from weight availability alone; no feedback loops (like issues/PRs) exist for model repositories \\

\hline
\end{tabular}
\caption{Indicative measurement signals for \emph{nexbax} dimensions. Existing benchmarks provide useful starting points for some dimensions, but many aspects of next-billion utility require proxy-based evidence and situated human evaluation.}
\label{tab:measurement_signals}
\end{table*}
As Table~\ref{tab:measurement_signals} illustrates, the measurement landscape is uneven. Some dimensions, such as efficiency, robustness, and multilingual performance, benefit from established benchmarks, while others—particularly adaptability, usability, and empowerment—remain under-specified in existing evaluation frameworks. This unevenness is not incidental but reflects the inherently context-dependent nature of utility in next-billion settings, motivating \emph{nexbax}'s hybrid measurement approach.

\subsection{Measurement Philosophy}

While \emph{nexbax} emphasizes measurable system properties, not all dimensions of utility can be fully captured through automated benchmarks or standardized metrics. In next-billion contexts, value is often emergent, context-dependent, and mediated by human judgment—particularly in environments characterized by infrastructural variability, informal workflows, and diverse user capabilities.

\emph{Nexbax} therefore adopts a hybrid measurement approach, combining:

\begin{itemize}[leftmargin=*,nolistsep]
    \item Objective signals, such as benchmark performance and observable system properties, and
    \item Situated evaluations, including interpretation by practitioners, public-sector implementers, civil society organizations, community intermediaries, and affected users grounded in real deployment contexts.
\end{itemize}

This reflects a fundamental asymmetry: while model capability can be evaluated in controlled environments, utility under constraint must be interpreted in context. Fully automated evaluation risks privileging what is easily measurable—such as latency, accuracy, or cost—over what is consequential but context-dependent, including ease of integration into informal workflows, cultural appropriateness, or accessibility for developers in low-resource environments. 
This issue is especially acute for cultural appropriateness. Recent work on cultural benchmarks argues that evaluation cannot be reduced to static question sets or language coverage alone: the questions, tasks, and interpretation criteria themselves must be developed with attention to the communities and contexts being represented \cite{chiu2024culturalbench}. \emph{Nexbax} therefore treats cultural fit as a participatory measurement problem, not merely a localization feature.

\emph{Nexbax} measures not only what AI systems are, but how they are experienced in use. 
We do not treat subjectivity as noise to be eliminated, but as signal that reflects variation in real-world applicability across settings. In practice, \emph{nexbax} complements local validation processes such as stakeholder consultation, domain-specific testing, institutional risk assessment, and deployment-specific evaluation rubrics. The index therefore functions as a comparative and diagnostic framework that in intended to be paired with context-sensitive interpretations of utility, risk, and institutional fit.

\subsection{Rubrics and Initial Scoring Exercise}

The measurement framework above defines what \emph{nexbax} seeks to capture in principle. To support comparative and formative assessment across diverse AI systems, we define broad rubrics that translate each dimension into a three-level scoring system (weak, moderate, strong). These rubrics simplify a multidimensional measurement space into interpretable evaluation criteria that can be applied consistently across systems and deployment settings. Rather than exhaustively capturing every aspect of a dimension, the rubrics 
provide a shared evaluative baseline that can be extended or adapted for particular organizational and domain needs.

Because usefulness depends on who is adopting or adapting a technology, the first version focuses on \textbf{developer-facing system properties}: how developer choices and capabilities shape the preconditions for AI utility and inclusion. This is an intentionally limited stakeholder layer, and future iterations should develop community- and institution-facing rubrics. We summarize the rubric in Table~\ref{tab:nexbax-rubrics}.
\begin{table*}[t]
\centering
\caption{Rubrics for the Next Billion AI Index}
\label{tab:nexbax-rubrics}

\renewcommand{\arraystretch}{1.3}

\begin{tabular}{p{0.18\textwidth} p{0.23\textwidth} p{0.23\textwidth} p{0.26\textwidth}}
\hline
\textbf{Dimension} & 
\textbf{1 -- weak} & 
\textbf{2 -- okay} & 
\textbf{3 -- strong} \\
\hline

\textbf{\textcolor{efficiency}{Cost-Effectiveness and \newline Price-Performance}} &
high TCO (usage cost + infra) for target users; no low-cost mode &
Some free/low-cost tier or decent on-device/edge efficiency &
clear low-cost path (free/community tier, metered API at local-friendly prices, runs on low-end devices/edge) \\
\hline

\textbf{\textcolor{efficiency}{Resource Efficiency}} &
heavy compute/infra, high latency, bandwidth hungry &
moderate footprint (e.g.\ via quantization/pruning); decent latency &
runs on low-spec edge/mobile; offline mode; optimized IO/latency \\
\hline

\textbf{\textcolor{practicality}{Adaptability and \newline Customizability}} &
little/no finetune or UI/config; one modality &
configurable; accepts adapters; multi-modal options &
modular and extensible: fine-tuning, plug-in tools, multi-modal, rapid pivots \\
\hline

\textbf{\textcolor{practicality}{Interoperability}} &
closed garden; limited standards/APIs &
some SDKs; limited adapters &
open standards plus rich connectors (tools, packages, webhooks); easy data in/out \\
\hline

\textbf{\textcolor{practicality}{Resilience, \newline Reliability, \newline and Robustness}} &
fails on flaky networks/dirty inputs; crashes; no retries &
survives short outages; tolerates moderate input noise; onus on infra provider &
designed for hostile environments: long offline queues/sync, power-safe modes, noise/dust tolerant pipelines, graceful degradation \\
\hline

\textbf{\textcolor{practicality}{Usability and \newline Automation}} &
expert-only; steep setup; code-only &
guided wizards/templates; some autoconfig; reasonable documentation &
no/low-code flows; opinionated templates; agentic automation; strong onboarding \\
\hline

\textbf{\textcolor{practicality}{Education and \newline Empowerment}} &
sparse documentation; no local-language help &
documentation and videos; community forum; some localization &
tutorials, local-language/ contextual help, community programs and NGO partners \\
\hline

\textbf{\textcolor{integrity}{Trustworthiness and Ethics}} &
opaque systems; unknown data; weak safety; no bias evaluation &
some documentation; basic filters; incident channel &
model/system cards; bias and safety evaluations across locales; user controls \\
\hline

\textbf{\textcolor{integrity}{Multiculturalism, \newline Inclusivity, and \newline Pluralism}} &
English-only; Western defaults &
major languages; partial localization; limited review of cultural assumptions &
broad language and dialect support; documented cultural assumptions; local stakeholder review; low-literacy UX and adaptation pathways \\
\hline

\textbf{\textcolor{integrity}{Openness and \newline Collaboration}} &
no weights or code; no roadmap or community &
some APIs or open SDKs; limited collaboration &
open weights or strong open standards; active community and public roadmap \\
\hline
\end{tabular}
\end{table*}
Given the hybrid measurement philosophy, the initial use of \emph{nexbax} relies on expert-driven formative assessment rather than fully automated scoring. Practitioners interpret rubric criteria in relation to their own deployment contexts. This first exercise assesses the clarity and perceived relevance of the index for technical and product stakeholders; it does not by itself validate usefulness for end users or communities.

\section{Formative Study}
\label{sec:evals}

For the formative exercise, we applied the rubrics in Table~\ref{tab:nexbax-rubrics} to three common AI stack configurations: open-weight models, closed models, and model-plus-orchestration/application layer systems. We selected these categories because they give practitioners a recognizable basis for discussion across architecture and governance choices. They are not intended to define the full scope of \emph{nexbax}; the same diagnostic framework can be adapted to other AI technologies, sectors, and deployment arrangements as the evidence base matures.

\subsection{Methods}
We conducted a formative expert evaluation of the first version of the index through one-hour, semi-structured user interviews with eleven subject matter experts working on AI technology catered to next billion markets, including founders, developers, and product leaders. The goal of this evaluation was to assess the clarity, actionability, and perceived accuracy of the dimensions for technical and product stakeholders, and to iterate on them based on expert feedback. Participants tested the index by rating three illustrative technology configurations (open weight models, closed weight models, and open weight model + orchestration layer) across the ten dimensions using the rubrics in Table~\ref{tab:nexbax-rubrics}. They shared feedback on their experience using the dimensions and rubrics, including where they saw value, points of confusion, and suggestions for improvement. While this sample is appropriate for probing developer-facing system properties, we note that it does not represent the full set of stakeholders needed to assess usefulness or cultural appropriateness in next-billion contexts. We summarize participants’ job roles, use cases, and geographies in Table~\ref{tab:evaluation-participants}.
\begin{table}[H]
\caption{Participants of the formative expert evaluation.}
\centering
\resizebox{\columnwidth}{!}{%
\begin{tabular}{c l l l}
\hline
\textbf{\#} & \textbf{Job Role} & \textbf{Use Case} & \textbf{Geography} \\
\hline
1  & Founder              & Visual analytics         & India \\
2  & Founder              & Content authentication   & India \\
3  & Engineering lead     & Content authentication   & India \\
4  & Head of AI           & Education                & India \\
5  & Operations \& analytics & Education            & India \\
6  & Product manager      & Customer service         & United States/India \\
7  & Founder \& CTO       & Visual analytics         & India \\
8  & Developer            & Information technology   & Kenya \\
9  & Engineering lead     & Information technology   & Kenya \\
10 & Data scientist       & Information technology   & Kenya \\
11 & CTO                  & AgTech                   & Ghana \\
\hline
\end{tabular}%
}
\label{tab:evaluation-participants}
\end{table}
\subsection{Results}
The outcomes of this evaluation were (1) formative support for the index’s perceived value and ease of use among technical stakeholders, and (2) refinements to the dimensions and rubrics. Overall, the experts we interviewed felt that the index was useful for reasoning about AI utility and adoption in next-billion settings from a developer and product perspective. As one participant explained, \emph{``all the tools that we use to evaluate our customers beyond localization, you hit all of them, like ease of use, reliability, the cost, the set up time — all those things are hit. Those are the same types of things that we use to pick our vendors.''} In particular, our findings indicate that cost effectiveness, usability \& automation, and trustworthiness \& ethics were the top three considerations for this set of experts. When asked to identify the most important decision-making considerations in their work, ten participants chose cost effectiveness, five chose usability \& automation, and four chose trustworthiness \& ethics. It is worth noting that these dimensions span all three themes, providing early signal that each theme captures distinct and meaningful aspects of real-world decision making. At the same time, specific dimensions will require different prioritization across markets and domains, and emphasizing one may come with trade-offs for another.

Participants also saw tangible value for their own work. For example, one participant envisioned the ratings serving as an \emph{``elimination mechanism''} to rule out technologies that don’t score well across dimensions; another speculated that \emph{``this framework would help me choose the right models to use.''} Similarly, one product leader described how the index would simplify their decision-making process:

\begin{quote}
\emph{``This kind of gives a summary or a digest...If you don't have something like this that...has already done the research for you, right? And it's presenting to you various options...you would have to sit down and do all of these things by yourself.''}
\end{quote}

A developer similarly spoke to the value of seeing expert evaluations based on the index:

\begin{quote}
\emph{``Within the community, I think there's no consensus on what works for us. So we don't have our models, so that's for sure, but then in the plethora of models that are accessible to us, which ones actually work for us? ...[The index] would be for us to, as a community, figure out what's best for us beyond just building policy documents, which is what countries have been doing. I think I would be very keen to see what even more experienced people, who've done a lot of work evaluating this, what they think.''}
\end{quote}

At the same time, a few participants felt that the index was limited in actionability and shared suggestions for improvement. Some felt that the dimensions were too high level for their specific use cases, and others felt that a numerical rating would be insufficient to capture nuances within a dimension. For example, within cost effectiveness, one participant differentiated between upfront cost (e.g., setting up a local service) vs. long-term costs (e.g., price hikes over time). Another participant gave similar examples: \emph{``education, I mean, is it materials for me as a dev [or] for me as...a non-tech user? ...Say, like, domain...when you talk about resource efficiency, there could be some groupings.''} To address these concerns, we recommend including explanations with ratings, as well as indicating the domain in which evaluations were conducted to support consumers in identifying evaluations relevant to them. As one participant suggested,

\begin{quote}
\emph{``It will be very useful for sure, but I would find myself trying to go beyond this and looking for some more information. Let's say, resource efficiency...it's rated as a two or three but why? And for my case, in that context...what does it mean at three?''}
\end{quote}

From a usability standpoint, the experts generally found the dimensions and rubrics easy to apply, although some felt that it would have been difficult without the guidance and clarifications they received from the researchers during the study session. In response to participants’ confusions, we made several clarifications that we anticipate will facilitate independent evaluation: adding a definition for each dimension, providing examples for a wider variety of use cases (e.g., vision and video analytics), and clarifying unfamiliar or confusing terms within dimension names and rubrics. For example, after five participants confused ``robustness'' with output accuracy, we de-emphasized it from the dimension name, renaming it from ``Robustness (Resilience and Reliability)'' to ''Resilience, Reliability, and Robustness.''

Finally, since our goal was to evaluate the index itself rather than collect accurate ratings for the three illustrative configurations, we omit the ratings collected across participants. A future iteration of this work will include aggregated, system-specific ratings from a larger and more diverse stakeholder sample, including public-sector implementers, civil society organizations, local domain experts, and affected users.

\section{Limitations and Future Work}
\label{sec:limitations}

\emph{Nexbax} should not be read as a universal measure of usefulness for the global majority. The present framework prioritizes system-level and developer-facing signals, such as affordability, deployment requirements, documentation, interoperability, and openness. These signals are important because they shape what kinds of AI systems can plausibly be adapted and sustained under constraints, but cannot determine whether a system is locally useful, culturally appropriate, or socially beneficial. Accordingly, the measurement signals in Table~\ref{tab:measurement_signals} should be read as indicative rather than definitive.

The formative evaluation is similarly limited. Participants brought relevant technical and product expertise, but they were primarily founders, developers, and product leaders. They therefore represent only one fraction of the stakeholder landscape for next-billion AI deployment. Future evaluation must include government officials, public-sector implementers, civil society organizations, community-based groups, domain workers, and affected users who can speak to local practices, institutional realities, and cultural meanings that technical experts may not observe.

Cross-context comparison also creates an unavoidable risk of flattening difference across countries, regions, languages, classes, and communities. \emph{Nexbax} should therefore be used with explanatory evidence and deployment context, not as a standalone scalar ranking. Future work will refine the mapping between dimensions, indicators, and scoring evidence through participatory, locale-specific evaluation protocols in which local stakeholders help define the questions, evidence standards, and interpretation criteria used to assess cultural appropriateness and practical usefulness.

More fundamentally, we do not claim to provide a complete or universal rubric for assessing usefulness across the extremely diverse populations, cultures, and institutional realities that make up the global majority. Rather, we see \emph{nexbax} as the beginning of a broader, grassroots effort to give greater voice and evaluative agency to stakeholders who are often underrepresented in dominant AI conversations. We therefore encourage researchers, practitioners, governments, and community organizations to adapt, extend, and reinterpret the framework for their own domains, deployment settings, and local priorities. The long-term value of \emph{nexbax} will depend not on enforcing a single canonical definition of useful AI, but on enabling locally grounded and participatory approaches to defining usefulness itself.

\section{Conclusion}
\label{sec:conclusions}

This paper began from a simple motivation: the systems most celebrated by current AI benchmarks are not necessarily the systems most likely to be useful for the next billion users. In next-billion contexts, adoption is shaped by affordability, unreliable infrastructure, linguistic diversity, informal and small-enterprise workflows, public-sector delivery channels, and uneven institutional capacity. Under these conditions, progress cannot be measured only through peak model capability. It must also be measured through the practical conditions that allow AI to be deployed, adapted, trusted, and sustained.

\emph{Nexbax} responds to this measurement gap by offering a first-stage diagnostic for artificial \emph{useful} intelligence. It brings together 10 dimensions under three themes: Effective Efficiency, Operational Practicality, and Societal Integrity. In doing so, the index translates broad goals of inclusive AI into concrete system-level properties, including cost-effectiveness, resource efficiency, adaptability, interoperability, robustness, usability and automation, education and empowerment, trustworthiness and ethics, multicultural alignment, and openness and collaboration. The accompanying rubrics make these properties discussable and comparable across models, orchestration frameworks, and application-layer systems, while preserving the central claim that next-billion usefulness depends on more than benchmark accuracy or model scale.

Our formative expert evaluation provides early support for this framing. Across eleven semi-structured interviews with founders, developers, product leaders, and technical practitioners working in next-billion AI markets, participants generally found the dimensions and rubrics relevant and useful for reasoning about technology choice, adoption trade-offs, and system fit. Their feedback also clarified where the framework can become more actionable: ratings should be accompanied by explanations, evaluations should specify deployment domain and context, and future versions should support more granular evidence within each dimension.

In that sense, the value of \emph{nexbax} is not only that it produces an index, but that it shifts the object of evaluation itself: from asking which AI systems are most capable in idealized settings to asking which systems create the conditions for useful, inclusive, and sustainable adoption where the next billion users live and work.

\clearpage

\section{Positionality Statement}

The authors of this paper bring interdisciplinary and cross-regional perspectives shaped by work in AI systems, HCI, trustworthy AI, product development, and AI governance across Africa, Asia, Europe, North America, and South America. Several authors have worked closely with communities and organizations operating in low-resource and next-billion-user contexts, including settings characterized by infrastructural constraints, linguistic diversity, and unequal access to digital technologies. These experiences informed the paper's focus on affordability, deployability, inclusivity, and contextual utility as central dimensions of AI adoption.

At the same time, the authors recognize that no single research team can fully represent the diversity of experiences and priorities across the global majority. The proposed framework is therefore intended not as a universal definition of useful AI, but as a starting point for participatory, context-sensitive evaluation and dialogue with local stakeholders and affected communities.

\section{Acknowledgments}
We thank Varun Aggarwal, Anthony Annunziata, Ginette Azcona, Daksh Chawla, Nahuel Defosse, Giridhar Ganapavarapu, George K. Githae, Sudeep Gowrishankar, Raghav Gupta, Jaikrishnan Hari, Ronak Khandelwal, Jordan McAfoose, Rohan Sahu, Siddhant Sachdeva, Het Shah, and Amith Singhee for their assistance with the survey process, feedback, and other contributions during the development of this work. Subhabrata Majumdar's research is supported by the Indian Institute of Management Bangalore Young Faculty Research Grant.

\bibliography{references}

\begin{thebibliography}{70}
\providecommand{\natexlab}[1]{#1}

\bibitem[{nat(2025)}]{nature2025localizing}
 2025.
\newblock Localizing AI in the global south.
\newblock \emph{Nature Machine Intelligence}, 7: 675.
\newblock Editorial.

\bibitem[{Achiam et~al.(2023)Achiam, Adler, Agarwal, Ahmad, Akkaya, Aleman, Almeida, Altenschmidt, Altman, Anadkat et~al.}]{achiam2023gpt}
Achiam, J.; Adler, S.; Agarwal, S.; Ahmad, L.; Akkaya, I.; Aleman, F.~L.; Almeida, D.; Altenschmidt, J.; Altman, S.; Anadkat, S.; et~al. 2023.
\newblock Gpt-4 technical report.
\newblock \emph{arXiv preprint arXiv:2303.08774}.

\bibitem[{Adams et~al.(2024)Adams, Adeleke, Florido, de~Magalhães~Santos, Grossman, Junck, and Stone}]{GIRAI2024}
Adams, R.; Adeleke, F.; Florido, A.; de~Magalhães~Santos, L.~G.; Grossman, N.; Junck, L.; and Stone, K. 2024.
\newblock Global Index on Responsible AI 2024 (1st Edition).
\newblock Technical report, South Africa: Global Center on AI Governance.

\bibitem[{Adler et~al.(2024)Adler, Agarwal, Aithal, Anh, Bhattacharya, Brundyn, Casper, Catanzaro, Clay, Cohen et~al.}]{adler2024nemotron}
Adler, B.; Agarwal, N.; Aithal, A.; Anh, D.~H.; Bhattacharya, P.; Brundyn, A.; Casper, J.; Catanzaro, B.; Clay, S.; Cohen, J.; et~al. 2024.
\newblock Nemotron-4 340b technical report.
\newblock \emph{arXiv preprint arXiv:2406.11704}.

\bibitem[{{African Union}(2024)}]{AU2024}
{African Union}. 2024.
\newblock Continental Artificial Intelligence Strategy.
\newblock \url{https://au.int/en/documents/20240809/continental-artificial-intelligence-strategy}.

\bibitem[{Anthropic(2024)}]{anthropic2024claude3}
Anthropic. 2024.
\newblock The Claude 3 model family: Opus, Sonnet, Haiku.
\newblock Technical report, Anthropic.

\bibitem[{{Anthropic}(2026)}]{Anthropic2026}
{Anthropic}. 2026.
\newblock Anthropic Economic Index.
\newblock \url{https://www.anthropic.com/research/economic-index-primitives}.

\bibitem[{Bailey et~al.(2026)Bailey, Kalarikalayil~Raju, Pearson, Robinson, and Jones}]{bailey2026street}
Bailey, G.; Kalarikalayil~Raju, D.; Pearson, J.; Robinson, S.; and Jones, M. 2026.
\newblock Street Scenes: Public Appliances for GenAI Video in Informal Settlements.
\newblock In \emph{Proceedings of the 2026 CHI Conference on Human Factors in Computing Systems}, 1--22.

\bibitem[{Banerjee and Duflo(2009)}]{BanerjeeDuflo2009}
Banerjee, A.~V.; and Duflo, E. 2009.
\newblock The experimental approach to development economics.
\newblock \emph{Annu. Rev. Econ.}, 1(1): 151--178.

\bibitem[{Banerjee and Duflo(2026)}]{banerjee2026poor}
Banerjee, A.~V.; and Duflo, E. 2026.
\newblock \emph{Poor economics: Rethinking poverty \& the ways to end it}.
\newblock Penguin Random House India Private Limited.

\bibitem[{Belcak et~al.(2025)Belcak, Heinrich, Diao, Fu, Dong, Muralidharan, Lin, and Molchanov}]{belcak2025small}
Belcak, P.; Heinrich, G.; Diao, S.; Fu, Y.; Dong, X.; Muralidharan, S.; Lin, Y.~C.; and Molchanov, P. 2025.
\newblock Small language models are the future of agentic ai.
\newblock \emph{arXiv preprint arXiv:2506.02153}.

\bibitem[{Borning and Muller(2012)}]{BorningMuller2012}
Borning, A.; and Muller, M. 2012.
\newblock Next Steps for Value Sensitive Design.
\newblock In \emph{Proceedings of the SIGCHI Conference on Human Factors in Computing Systems}.

\bibitem[{Casper et~al.(2025)Casper, O'Brien, Longpre, Seger, Klyman, Bommasani, Nrusimha, Shumailov, Mindermann, Basart et~al.}]{casper2025open}
Casper, S.; O'Brien, K.; Longpre, S.; Seger, E.; Klyman, K.; Bommasani, R.; Nrusimha, A.; Shumailov, I.; Mindermann, S.; Basart, S.; et~al. 2025.
\newblock Open technical problems in open-weight AI model risk management.
\newblock \emph{Transactions on Machine Learning Research}.

\bibitem[{Chandiramani et~al.(2026)Chandiramani, Blakeman, Olaoye, Gupta, Somasamudramath, Khattar, Adesoba, Renduchintala, Asif, Agrawal et~al.}]{chandiramani2026nemotron}
Chandiramani, A.; Blakeman, A.; Olaoye, A.; Gupta, A.; Somasamudramath, A.; Khattar, A.; Adesoba, A.; Renduchintala, A.; Asif, A.; Agrawal, A.; et~al. 2026.
\newblock Nemotron 3 Super: Open, Efficient Mixture-of-Experts Hybrid Mamba-Transformer Model for Agentic Reasoning.
\newblock \emph{arXiv preprint arXiv:2604.12374}.

\bibitem[{Chiu et~al.(2024)Chiu, Jiang, Lin, Park, Li, Ravi, Bhatia, Antoniak, Tsvetkov, Shwartz et~al.}]{chiu2024culturalbench}
Chiu, Y.~Y.; Jiang, L.; Lin, B.~Y.; Park, C.~Y.; Li, S.~S.; Ravi, S.; Bhatia, M.; Antoniak, M.; Tsvetkov, Y.; Shwartz, V.; et~al. 2024.
\newblock CulturalBench: a Robust, Diverse and Challenging Benchmark on Measuring (the Lack of) Cultural Knowledge of LLMs.

\bibitem[{Clancy, Zhu, and Majumdar(2025)}]{f2025exploring}
Clancy, R.~F.; Zhu, Q.; and Majumdar, S. 2025.
\newblock Exploring AI ethics in global contexts: a culturally responsive, psychologically realist approach.
\newblock \emph{AI and Ethics}, 5(6): 6329--6338.

\bibitem[{Eiras et~al.(2024)Eiras, Petrov, Vidgen, Schroeder, Pizzati, Elkins, Mukhopadhyay, Bibi, Purewal, Botos et~al.}]{eiras2024risks}
Eiras, F.; Petrov, A.; Vidgen, B.; Schroeder, C.; Pizzati, F.; Elkins, K.; Mukhopadhyay, S.; Bibi, A.; Purewal, A.; Botos, C.; et~al. 2024.
\newblock Risks and opportunities of open-source generative ai.
\newblock \emph{arXiv preprint arXiv:2405.08597}.

\bibitem[{Enevoldsen et~al.(2025)Enevoldsen, Chung, Kerboua, Kardos, Mathur, Stap, Gala, Siblini, Krzemi{\'n}ski, Winata et~al.}]{enevoldsen2025mmteb}
Enevoldsen, K.; Chung, I.; Kerboua, I.; Kardos, M.; Mathur, A.; Stap, D.; Gala, J.; Siblini, W.; Krzemi{\'n}ski, D.; Winata, G.~I.; et~al. 2025.
\newblock Mmteb: Massive multilingual text embedding benchmark.
\newblock \emph{arXiv preprint arXiv:2502.13595}.

\bibitem[{FMCIDE(2025)}]{NigeriaNAIS2025}
FMCIDE. 2025.
\newblock National Artificial Intelligence Strategy (2025--2029).
\newblock \url{https://fmcide.gov.ng/initiative/nais/}.

\bibitem[{Ghosh et~al.(2025)Ghosh, Frase, Williams, Luger, R{\"o}ttger, Barez, McGregor, Fricklas, Kumar, Bollacker et~al.}]{ghosh2025ailuminate}
Ghosh, S.; Frase, H.; Williams, A.; Luger, S.; R{\"o}ttger, P.; Barez, F.; McGregor, S.; Fricklas, K.; Kumar, M.; Bollacker, K.; et~al. 2025.
\newblock Ailuminate: Introducing v1. 0 of the ai risk and reliability benchmark from mlcommons.
\newblock \emph{arXiv preprint arXiv:2503.05731}.

\bibitem[{Gibbons(2018)}]{gibbons2018empathy}
Gibbons, S. 2018.
\newblock Empathy Mapping: The First Step in Design Thinking.
\newblock Nielsen Norman Group.

\bibitem[{{Government of India}(2026{\natexlab{a}})}]{indiaai2026impact}
{Government of India}. 2026{\natexlab{a}}.
\newblock AI Impact Summit 2026.
\newblock \url{https://impact.indiaai.gov.in/}.

\bibitem[{{Government of India}(2026{\natexlab{b}})}]{IndiaAI}
{Government of India}. 2026{\natexlab{b}}.
\newblock IndiaAI Mission.
\newblock \url{https://indiaai.gov.in/}.

\bibitem[{{Government of Kenya}(2025)}]{KenyaAI2025}
{Government of Kenya}. 2025.
\newblock Kenya Artificial Intelligence Strategy 2025--2030.

\bibitem[{Goyal et~al.(2022)Goyal, Gao, Chaudhary, Chen, Wenzek, Ju, Krishnan, Ranzato, Guzm{\'a}n, and Fan}]{goyal2022flores}
Goyal, N.; Gao, C.; Chaudhary, V.; Chen, P.-J.; Wenzek, G.; Ju, D.; Krishnan, S.; Ranzato, M.; Guzm{\'a}n, F.; and Fan, A. 2022.
\newblock The Flores-101 evaluation benchmark for low-resource and multilingual machine translation.
\newblock \emph{Transactions of the Association for Computational Linguistics}, 10: 522--538.

\bibitem[{{ISO and IEC}(2023)}]{ISOIEC420012023}
{ISO and IEC}. 2023.
\newblock {ISO/IEC 42001:2023 Information Technology -- Artificial Intelligence -- Management System}.
\newblock \url{https://www.iso.org/standard/42001}.

\bibitem[{Jiang et~al.(2023)Jiang, Sablayrolles, Mensch, Bamford, Chaplot, de~las Casas, Bressand, Lengyel, Lample, Saulnier et~al.}]{jiang2023mistral}
Jiang, A.~Q.; Sablayrolles, A.; Mensch, A.; Bamford, C.; Chaplot, D.~S.; de~las Casas, D.; Bressand, F.; Lengyel, G.; Lample, G.; Saulnier, L.; et~al. 2023.
\newblock Mistral 7B.
\newblock \emph{arXiv preprint arXiv:2310.06825}.

\bibitem[{Jiang et~al.(2024)Jiang, Sablayrolles, Roux, Mensch, Lavall{\'e}e, Douillard, El-Sayed, Flagel, Galtier, Gombar et~al.}]{jiang2024mixtral}
Jiang, A.~Q.; Sablayrolles, A.; Roux, A.; Mensch, A.; Lavall{\'e}e, B.; Douillard, X.; El-Sayed, W.; Flagel, A.; Galtier, D.; Gombar, J.; et~al. 2024.
\newblock Mixtral of Experts.
\newblock \emph{arXiv preprint arXiv:2401.04088}.

\bibitem[{Jimenez et~al.(2024)Jimenez, Yang, Wettig, Yao, Pei, Press, and Narasimhan}]{jimenez2024swe}
Jimenez, C.~E.; Yang, J.; Wettig, A.; Yao, S.; Pei, K.; Press, O.; and Narasimhan, K. 2024.
\newblock Swe-bench: Can language models resolve real-world github issues?
\newblock In \emph{International Conference on Learning Representations}, volume 2024, 54107--54157.

\bibitem[{Karnani(2007)}]{Karnani2007}
Karnani, A. 2007.
\newblock The mirage of marketing to the bottom of the pyramid: How the private sector can help alleviate poverty.
\newblock \emph{California management review}, 49(4): 90--111.

\bibitem[{Kim et~al.(2025)Kim, Kuehnert, Lazar, Singh, and Heidari}]{kim2025ai}
Kim, R.~M.; Kuehnert, B.; Lazar, S.; Singh, R.; and Heidari, H. 2025.
\newblock The AI Power Disparity Index: Toward a Compound Measure of AI Actors’ Power to Shape the AI Ecosystem.
\newblock In \emph{Proceedings of the AAAI/ACM Conference on AI, Ethics, and Society}, volume~8, 1453--1464.

\bibitem[{Kirk et~al.(2024)Kirk, Whitefield, R{\"o}ttger, Bean, Margatina, Ciro, Mosquera, Bartolo, Williams, He et~al.}]{kirk2024prism}
Kirk, H.~R.; Whitefield, A.; R{\"o}ttger, P.; Bean, A.; Margatina, K.; Ciro, J.; Mosquera, R.; Bartolo, M.; Williams, A.; He, H.; et~al. 2024.
\newblock The PRISM alignment dataset: What participatory, representative and individualised human feedback reveals about the subjective and multicultural alignment of large language models.
\newblock \emph{Advances in Neural Information Processing Systems}, 37: 105236--105344.

\bibitem[{{KPMG}(2025)}]{KPMG2025}
{KPMG}. 2025.
\newblock Trust, Attitudes and Use of Artificial Intelligence.
\newblock \url{https://kpmg.com/xx/en/our-insights/ai-and-technology/trust-attitudes-and-use-of-ai.html}.

\bibitem[{Liang et~al.(2023)Liang, Bommasani, Lee, Tsipras, Soylu, Yasunaga, Zhang, Narayanan, Wu, Kumar, Newman, Yuan, Yan, Zhang, Cosgrove, Manning, Re, Acosta-Navas, Hudson, Zelikman, Durmus, Ladhak, Rong, Ren, Yao, Wang, Santhanam, Orr, Zheng, Yuksekgonul, Suzgun, Kim, Guha, Chatterji, Khattab, Henderson, Huang, Chi, Xie, Santurkar, Ganguli, Hashimoto, Icard, Zhang, Chaudhary, Wang, Li, Mai, Zhang, and Koreeda}]{liang2023holistic}
Liang, P.; Bommasani, R.; Lee, T.; Tsipras, D.; Soylu, D.; Yasunaga, M.; Zhang, Y.; Narayanan, D.; Wu, Y.; Kumar, A.; Newman, B.; Yuan, B.; Yan, B.; Zhang, C.; Cosgrove, C.~A.; Manning, C.~D.; Re, C.; Acosta-Navas, D.; Hudson, D.~A.; Zelikman, E.; Durmus, E.; Ladhak, F.; Rong, F.; Ren, H.; Yao, H.; Wang, J.; Santhanam, K.; Orr, L.; Zheng, L.; Yuksekgonul, M.; Suzgun, M.; Kim, N.; Guha, N.; Chatterji, N.~S.; Khattab, O.; Henderson, P.; Huang, Q.; Chi, R.~A.; Xie, S.~M.; Santurkar, S.; Ganguli, S.; Hashimoto, T.; Icard, T.; Zhang, T.; Chaudhary, V.; Wang, W.; Li, X.; Mai, Y.; Zhang, Y.; and Koreeda, Y. 2023.
\newblock Holistic Evaluation of Language Models.
\newblock \emph{Transactions on Machine Learning Research}.

\bibitem[{Luccioni et~al.(2024)Luccioni, Gamazaychikov, Strubell, Hooker, Jernite, Mitchell, and Wu}]{AIEnergyScore2026}
Luccioni, S.; Gamazaychikov, B.; Strubell, E.; Hooker, S.; Jernite, Y.; Mitchell, M.; and Wu, C.-J. 2024.
\newblock AI Energy Score.
\newblock \url{https://huggingface.co/AIEnergyScore}.

\bibitem[{Melville et~al.(2013)Melville, Chenthamarakshan, Lawrence, Powell, Mugisha, Sapra, Anandan, and Assefa}]{melville2013amplifying}
Melville, P.; Chenthamarakshan, V.; Lawrence, R.~D.; Powell, J.; Mugisha, M.; Sapra, S.; Anandan, R.; and Assefa, S. 2013.
\newblock Amplifying the voice of youth in africa via text analytics.
\newblock In \emph{Proceedings of the 19th ACM SIGKDD international conference on Knowledge discovery and data mining}, 1204--1212.

\bibitem[{{METR}(2025)}]{METRAutonomousCapabilities}
{METR}. 2025.
\newblock Resources for Measuring Autonomous AI Capabilities.
\newblock \url{https://metr.org/measuring-autonomous-ai-capabilities/}.

\bibitem[{{METR}(2026)}]{METRTimeHorizons}
{METR}. 2026.
\newblock Task-Completion Time Horizons of Frontier AI Models.
\newblock \url{https://metr.org/time-horizons/}.

\bibitem[{{Microsoft}(2025)}]{Microsoft2025}
{Microsoft}. 2025.
\newblock Microsoft New Future of Work 2025.
\newblock \url{https://www.microsoft.com/en-us/research/project/the-new-future-of-work/}.

\bibitem[{{Microsoft Research Africa}(2025)}]{atlas2025}
{Microsoft Research Africa}. 2025.
\newblock Atlas: A Playbook for Cross Cultural AI.
\newblock \url{https://github.com/microsoft/Atlas}.
\newblock A practical guide conducting cross-cultural AI Research for Development and Deployment.

\bibitem[{Microsoft Research~Africa(2026)}]{pazabench2026}
Microsoft Research~Africa, N. 2026.
\newblock PazaBench: A Benchmark for Automatic Speech Recognition on Low Resource Languages.
\newblock \url{https://www.microsoft.com/en-us/research/project/project-gecko/}.
\newblock Alpha version. Part of Project Gecko - Equitable Generative AI for the Global Majority.

\bibitem[{Miranda et~al.(2026)Miranda, Hu, Reichart, and Korhonen}]{miranda2026multilinguality}
Miranda, L. J.~V.; Hu, S.; Reichart, R.; and Korhonen, A. 2026.
\newblock Multilinguality at the Edge: Developing Language Models for the Global South.
\newblock \emph{arXiv preprint arXiv:2604.21637}.

\bibitem[{{MIT Media Lab}(2026)}]{ImpactBench2026}
{MIT Media Lab}. 2026.
\newblock ImpactBench.
\newblock \url{https://impactbench.media.mit.edu/}.

\bibitem[{Mohamed, Png, and Isaac(2020)}]{Mohamed2020}
Mohamed, S.; Png, M.-T.; and Isaac, W. 2020.
\newblock Decolonial AI: Decolonial Theory as Sociotechnical Foresight in Artificial Intelligence.
\newblock \emph{Philosophy \& Technology}, 33(4): 659--684.

\bibitem[{Muchai et~al.(2026)Muchai, Chege, Mumero, and Nyairo}]{muchai2026paza}
Muchai, M.; Chege, K.; Mumero, N.; and Nyairo, S. 2026.
\newblock Paza: Introducing automatic speech recognition benchmarks and models for low-resource languages.
\newblock \url{https://www.microsoft.com/en-us/research/blog/paza-introducing-automatic-speech-recognition-benchmarks-and-models-for-low-resource-languages/}.

\bibitem[{Muennighoff et~al.(2023)Muennighoff, Tazi, Magne, and Reimers}]{muennighoff2023mteb}
Muennighoff, N.; Tazi, N.; Magne, L.; and Reimers, N. 2023.
\newblock Mteb: Massive text embedding benchmark.
\newblock In \emph{Proceedings of the 17th Conference of the European Chapter of the Association for Computational Linguistics}, 2014--2037.

\bibitem[{{NITI Aayog}(2025)}]{NITIAayog2025}
{NITI Aayog}. 2025.
\newblock AI for Viksit Bharat: The Opportunity of Accelerated Economic Growth.
\newblock \url{https://niti.gov.in/whats-new/ai-viksit-bharat-opportunity-accelerated-economic-growth}.

\bibitem[{Odame-Darkwah and Tsado(2025)}]{odame-darkwah2025isf}
Odame-Darkwah, K.; and Tsado, A. 2025.
\newblock ISF Voices 2025: Africa’s Playbook.
\newblock Substack article.

\bibitem[{{OECD}(2026)}]{OECD2026}
{OECD}. 2026.
\newblock The OECD.AI Index.
\newblock \url{https://doi.org/10.1787/32c01014-en}.

\bibitem[{Oluwatuyi et~al.(2026)Oluwatuyi, Pillay, Mazwi, Castro, Farao, Srivatsa, Bagalkot, and Densmore}]{oluwatuyi2026collectively}
Oluwatuyi, R.; Pillay, V.; Mazwi, J.; Castro, A.; Farao, J.; Srivatsa, S.; Bagalkot, N.; and Densmore, M. 2026.
\newblock Collectively Reimagining Artificial Intelligence With Marginalized Communities.
\newblock In \emph{Proceedings of the 2026 CHI Conference on Human Factors in Computing Systems}, 1--21.

\bibitem[{{Oxford Insights}(2025)}]{Oxford2025}
{Oxford Insights}. 2025.
\newblock Government AI Readiness Index.
\newblock \url{https://oxfordinsights.com/ai-readiness/}.

\bibitem[{Prabhakaran et~al.(2022)Prabhakaran, Mitchell, Gebru, and Gabriel}]{Prabhakaran2022}
Prabhakaran, V.; Mitchell, M.; Gebru, T.; and Gabriel, I. 2022.
\newblock A Human Rights-Based Approach to Responsible AI.
\newblock \emph{arXiv preprint arXiv:2210.02667}.

\bibitem[{Prahalad(2006)}]{Prahalad2006}
Prahalad, C.~K. 2006.
\newblock \emph{The Fortune at the Bottom of the Pyramid: Eradicating Poverty Through Profits}.
\newblock Upper Saddle River, NJ: Prentice Hall.

\bibitem[{Prahalad(2012)}]{Prahalad2012}
Prahalad, C.~K. 2012.
\newblock Bottom of the Pyramid as a Source of Breakthrough Innovations.
\newblock \emph{Journal of Product Innovation Management}.

\bibitem[{Reddi et~al.(2020)Reddi, Cheng, Kanter, Mattson, Schmuelling, Wu, Anderson, Breughe, Charlebois, Chou et~al.}]{reddi2020mlperf}
Reddi, V.~J.; Cheng, C.; Kanter, D.; Mattson, P.; Schmuelling, G.; Wu, C.-J.; Anderson, B.; Breughe, M.; Charlebois, M.; Chou, W.; et~al. 2020.
\newblock MLPerf inference benchmark.
\newblock In \emph{2020 ACM/IEEE 47th Annual International Symposium on Computer Architecture (ISCA)}, 446--459. IEEE.

\bibitem[{Rismani et~al.(2025)Rismani, Shelby, Davis, Rostamzadeh, and Moon}]{Rismani_Shelby_Davis_Rostamzadeh_Moon_2025}
Rismani, S.; Shelby, R.; Davis, L.; Rostamzadeh, N.; and Moon, A. 2025.
\newblock Measuring What Matters: Connecting AI Ethics Evaluations to System Attributes, Hazards, and Harms.
\newblock \emph{Proceedings of the AAAI/ACM Conference on AI, Ethics, and Society}, 8(3): 2199–2213.

\bibitem[{Roberts(2024)}]{Roberts2024DigitalSovereignty}
Roberts, H. 2024.
\newblock Digital Sovereignty and Artificial Intelligence: A Normative Approach.
\newblock \emph{Ethics and Information Technology}, 26(70).

\bibitem[{Saad-Falcon et~al.(2025)Saad-Falcon, Narayan, Akengin, Griffin, Shandilya, Lafuente, Goel, Joseph, Natarajan, Guha et~al.}]{saad2025intelligence}
Saad-Falcon, J.; Narayan, A.; Akengin, H.~O.; Griffin, J.; Shandilya, H.; Lafuente, A.~G.; Goel, M.; Joseph, R.; Natarajan, S.; Guha, E.~K.; et~al. 2025.
\newblock Intelligence per watt: Measuring intelligence efficiency of local ai.
\newblock \emph{arXiv preprint arXiv:2511.07885}.

\bibitem[{Sambasivan et~al.(2021)Sambasivan, Arnesen, Hutchinson, Doshi, and Prabhakaran}]{Sambasivan2021}
Sambasivan, N.; Arnesen, E.; Hutchinson, B.; Doshi, T.; and Prabhakaran, V. 2021.
\newblock Re-imagining Algorithmic Fairness in India and Beyond.
\newblock \emph{Proceedings of the 2021 ACM Conference on Fairness, Accountability, and Transparency}.

\bibitem[{Seger et~al.(2023)Seger, Dreksler, Moulange, Dardaman, Schuett, Wei, Winter, Arnold, h{\'E}igeartaigh, Korinek et~al.}]{seger2023open}
Seger, E.; Dreksler, N.; Moulange, R.; Dardaman, E.; Schuett, J.; Wei, K.; Winter, C.; Arnold, M.; h{\'E}igeartaigh, S.~{\'O}.; Korinek, A.; et~al. 2023.
\newblock Open-sourcing highly capable foundation models: An evaluation of risks, benefits, and alternative methods for pursuing open-source objectives.
\newblock \emph{arXiv preprint arXiv:2311.09227}.

\bibitem[{Shoham et~al.(2017)Shoham, Perrault, Brynjolfsson, and Clark}]{AIIndex2017}
Shoham, Y.; Perrault, R.; Brynjolfsson, E.; and Clark, J. 2017.
\newblock Artificial Intelligence Index Report.
\newblock Technical report, Stanford University, Human-Centered Artificial Intelligence.

\bibitem[{Touvron et~al.(2023)Touvron, Lavril, Izacard, Martinet, Lachaux, Lacroix, Rozi{\`e}re, Goyal, Hambro, Azhar et~al.}]{touvron2023llama}
Touvron, H.; Lavril, T.; Izacard, G.; Martinet, X.; Lachaux, M.-A.; Lacroix, T.; Rozi{\`e}re, B.; Goyal, N.; Hambro, E.; Azhar, F.; et~al. 2023.
\newblock Llama: Open and efficient foundation language models.
\newblock \emph{arXiv preprint arXiv:2302.13971}.

\bibitem[{{UNDP}(2025)}]{UNDP2025}
{UNDP}. 2025.
\newblock Human Development Report 2025: A Matter of Choice: People and Possibilities in the Age of AI.
\newblock Technical report, UNDP, New York.

\bibitem[{{UNESCO}(2023)}]{UNESCO2023}
{UNESCO}. 2023.
\newblock Readiness Assessment Methodology: A Tool of the Recommendation on the Ethics of Artificial Intelligence.
\newblock Technical report, UNESCO.

\bibitem[{Varshney(2023)}]{varshney2023foundation}
Varshney, K.~R. 2023.
\newblock Foundation Model Platforms and Bottom-of-the-Pyramid Innovation.
\newblock In \emph{ICLR Workshop on Practical Machine Learning for Developing Countries}.

\bibitem[{Varuvel~Dennison et~al.(2026)Varuvel~Dennison, Jain, Ganu, and Vashistha}]{varuvel2026designing}
Varuvel~Dennison, D.; Jain, M.; Ganu, T.; and Vashistha, A. 2026.
\newblock Designing Culturally Aligned AI Systems For Social Good in Non-Western Contexts.
\newblock In \emph{Proceedings of the 2026 CHI Conference on Human Factors in Computing Systems}, 1--22.

\bibitem[{{Vibhasha Team}(2025)}]{microsoft2024vibhasha}
{Vibhasha Team}. 2025.
\newblock Vibhasha: The Multilingual Playbook for Large Language Models.
\newblock \url{https://github.com/microsoft/vibhasha-playbook}.
\newblock A practical guide for developing multilingual and culturally-aware LLM systems.

\bibitem[{Wan et~al.(2023)Wan, Kyleman, Kapoor, Maslej, Longpre, Xiong, Liang, and Bommasani}]{FMTI2023}
Wan, A.; Kyleman, K.; Kapoor, S.; Maslej, N.; Longpre, S.; Xiong, B.; Liang, P.; and Bommasani, R. 2023.
\newblock Foundation Model Transparency Index.

\bibitem[{Wang et~al.(2023)Wang, Chen, Pei, Xie, Kang, Zhang, Xu, Xiong, Dutta, Schaeffer et~al.}]{wang2023decodingtrust}
Wang, B.; Chen, W.; Pei, H.; Xie, C.; Kang, M.; Zhang, C.; Xu, C.; Xiong, Z.; Dutta, R.; Schaeffer, R.; et~al. 2023.
\newblock DecodingTrust: A Comprehensive Assessment of Trustworthiness in $\{$GPT$\}$ Models.
\newblock \emph{Neural Information Processing Systems Datasets; Benchmarks Track}.

\bibitem[{Zhuo et~al.(2025)Zhuo, Vu, Chim, Hu, Yu, Widyasari, Yusuf, Zhan, He, Paul et~al.}]{zhuo2025bigcodebench}
Zhuo, T.~Y.; Vu, M.~C.; Chim, J.; Hu, H.; Yu, W.; Widyasari, R.; Yusuf, I. N.~B.; Zhan, H.; He, J.; Paul, I.; et~al. 2025.
\newblock Bigcodebench: Benchmarking code generation with diverse function calls and complex instructions.
\newblock In \emph{International Conference on Learning Representations}, volume 2025, 66602--66656.

\end{thebibliography}


\end{document}